\documentclass[twocolumn,showpacs,showkeys,preprintnumbers,amsmath,amssymb,prd]{revtex4}

\usepackage{graphicx}
\usepackage{dcolumn}
\usepackage{bm}

\begin{document}


\title{Magnetic monopoles with generalized quantization condition}

\author{Alexander I  Nesterov}
\altaffiliation[Also at ]{L.V. Kirensky Institute of
Physics, Krasnoyarsk, Russia}
\email{nesterov@udgserv.cencar.udg.mx}
\homepage{http://udgserv.cencar.udg.mx/~nesterov}

\author{Ferm\'\i n Aceves de la Cruz}%
\altaffiliation[Also at ]{Instituto de F{\'\i}sica,
Universidad de Guanajuato, Le\'on, Guanajuato, M\'exico}
\email{fermin@udgphys.intranets.com}

\affiliation{Departamento de F{\'\i}sica, CUCEI, Universidad de
Guadalajara, Guadalajara, Jalisco, M\'exico }

\date{\today}

\begin{abstract}
Theory of pointlike magnetic monopole with an arbitrary magnetic
charge is considered. It is shown that a proper description
requires making use of nonunitary representations of the rotation
group and the nonassociative generalization of the gauge group and
fibre bundle theory.

\end{abstract}



\maketitle

\section{Introduction}

In his remarkable paper Dirac \cite{Dir} showed that a proper
description of the quantum mechanics of a charged particle of the
charge $e$ in the field of the magnetic monopole of the charge $q$
requires the quantization condition $2\mu \in \mathbb Z$ (we set
$\mu=eq$ and $\hbar = c =1$). There are strong mathematical and
physical arguments why this condition must be fulfilled
\cite{Dir,Sw_1,Wu1,Wu2,Jac,Gr,Gr1,G1,G2,G3}. For instance, it
restores associativity of the translation group for the
charge-monopole system, ensures the absence of an Aharanov-Bohm
effect produced by a Dirac string, arises as natural condition of
the description pointlike Abelian magnetic monopole in the
framework of fibre bundle theory. Finally, Dirac's quantization
condition can be derived employing the unitary representation of
the rotation group.

In our paper we show that there exists the consistent theory of
the magnetic monopole with an arbitrary magnetic charge. It
requires nonunitary representations of the rotation group and
nonassociative generalization of gauge transformations and fibre
bundles theory, where a gauge group is replaced by gauge loop.

\section{Preliminaries}

 A magnetic field of the monopole is
\begin{equation}
{\mathbf B} = q \frac{\mathbf r}{r^3}, \label{eq_0}
\end{equation}
and as well known any choice of the vector potential $\mathbf A$
being compatible with Eq. (\ref{eq_0}) must have singularities.
For instance, Dirac introduced the vector potential as
\begin{equation}
{\mathbf A}_{\mathbf n}= q\frac{{\mathbf r}\times {\mathbf n}}
{r(r - {\mathbf n} \cdot{\mathbf r})} \label{d_str}
\end{equation}
where the unit vector $\mathbf n$ determines the direction of a
string $S_{\mathbf n}$ passing from the origin of coordinates to
$\infty$ \cite{Dir}. Schwinger's choice is
\begin{equation}
{\mathbf A^{SW}}= \frac{1}{2}\bigl({\mathbf A}_{\mathbf n}+
{\mathbf A}_{-\mathbf n} \bigr) = q\frac{({\mathbf n}
\cdot{\mathbf r}){\mathbf r}\times {\mathbf n}} {r\bigl(r^2 -
({\mathbf n} \cdot{\mathbf r})^2\bigr)}, \label{sw}
\end{equation}
and the string is propagated from $-\infty$ to $\infty$
\cite{Sw_1}.

It is easy verify that
\[
{\rm rot}{\mathbf A}_{\mathbf n} ={\mathbf B} - {\mathbf
h}_{\mathbf n}, \quad {\rm rot}{\mathbf A}^{SW} ={\mathbf B} -
{\mathbf h}^{SW}
\]
where
\begin{eqnarray}
&&{\mathbf h}_{\mathbf n}  = 4\pi q{\mathbf n}\int _{0}^\infty
\delta^3(\mathbf r - \mathbf n \tau) d \tau ,\\
&&{\mathbf h}^{SW}  = 2\pi q{\mathbf n}\int _{-\infty}^\infty
\delta^3(\mathbf r - \mathbf n \tau) d \tau
\end{eqnarray}
determine the magnetic field of the respective strings. Both
vector potentials yield the same magnetic monopole field, however
the quantization is different, while the Dirac condition is
$2\mu=p$, the Schwinger one is $\mu=p, \; p\in \mathbb Z$.

These two strings are members of a family $\{{\mathbf
S}^{\kappa}_{\mathbf n} \}$ with the magnetic field given by
\begin{align}
&{\mathbf h}^{\kappa}_{\mathbf n}=  \kappa{\mathbf h}_{\mathbf n}
+ (1-\kappa){\mathbf h}_{-\mathbf n} \label{str}
\end{align}
where $\kappa$ is a weight of a semi-infinite Dirac's string.
Further we call ${\mathbf S}^{\kappa}_{\mathbf n}$ a {\it weighted
string}.

For a non relativistic charged particle in the field of a magnetic
monopole the equations of motion
\begin{equation}
\ddot{\mathbf r} = \frac{\mu}{r^3}{\mathbf r} \times\dot{\mathbf
r} \label{eq1}
\end{equation}
imply that the total angular momentum
\begin{equation}
{\mathbf J} = {\mathbf r} \times \left({\mathbf p} - e{\mathbf
A}\right) - \mu\frac{\mathbf r}{r} \label{eq1c}
\end{equation}
is conserved. The last term in Eq.(\ref{eq1c}) usually is
interpreted as the contribution of the electromagnetic field,
which carries an angular momentum \cite{Gol1,Gol2,Lyn}
\[
{\mathbf L}_{em}=\frac{1}{4\pi}\int {\mathbf r} \times({\mathbf
E}\times{\mathbf B}) d^3 r = - \mu\frac{\mathbf r}{r}.
\]

The operator
\begin{equation}
{\mathbf J} = {\mathbf r} \times \left({-i\mathbf \nabla} -
e{\mathbf A}\right) - \mu\frac{\mathbf r}{r},
\end{equation}
representing the angular momentum $\mathbf J$, has the same
properties as a standard angular momentum and obeys the following
commutation relations:
\begin{eqnarray}
&&[H, {\mathbf J}^2] = 0, \quad [H, J_i] = 0,\quad  [{\mathbf
J}^2, J_i] =
0, \label{eq5a} \\
&&[J_i, J_j] = i\epsilon_{ijk}J_k \label{eq5}
\end{eqnarray}
where $H$ is the Hamiltonian. Notice that the commutation
relations fail on the string, however, $H$ and $\mathbf J$ may be
extended to self-adjoint operators satisfying the commutation
relations of Eqs. (\ref{eq5a}), (\ref{eq5})  for any value of
$\mu$ \cite{Zw_1,H,Str}.

Now following \cite{Wu1,Wu2}, let us cover the two-dimensional
sphere $S^2$ of fixed radius $r>0$ by two neighborhoods $0 \leq
\theta < \pi/2 + \varepsilon$ and $\pi/2 -\varepsilon \leq \theta
< \pi$. The vector potential is taken to be
\begin{eqnarray}
{\mathbf A_N} =  q\frac{1-\cos{\theta}}{r\sin{\theta}}\;
\hat{\mathbf e}_{\varphi}, \quad {\mathbf A_S} =
-q\frac{1+\cos\theta}{r\sin\theta}\; \hat{\mathbf e}_{\varphi}
\label{eq1b}
\end{eqnarray}
where $(r,\theta,\varphi)$ are the spherical coordinates. Notice
that ${\mathbf A_{N,S}}$ have singularities on $(S,N)$ pole of the
sphere and in the overlap of the neghborhoods ${\mathbf A_N}$ and
${\mathbf A_S}$  are related by a gauge transformation.

Choosing the vector potential as $\mathbf A_N$ we have
\begin{eqnarray}
&&J_{\pm}= e^{\pm
i\varphi}\bigg(\pm\frac{\partial}{\partial\theta} +i\cot\theta
\frac{\partial}{\partial\varphi} -
\frac{\mu\sin\theta}{1+\cos\theta} \bigg),\\
&&J_0=-i\frac{\partial}{\partial\varphi} - \mu,\\
&&{\mathbf J^2} =-\frac{1}{\sin{\theta}}
\frac{\partial~}{\partial\theta}\left(\sin{\theta}
\frac{\partial~}{\partial\theta}\right) -
\frac{1}{\sin^2{\theta}}\frac{\partial^2~}{\partial\varphi^2} + \nonumber\\
&&+i\frac{2\mu}{1 +\cos{\theta}}\frac{\partial~}{\partial\varphi}
+\mu^2\frac{1 - \cos{\theta}}{1 + \cos{\theta}} +\mu^2 \label{eq7}
\end{eqnarray}
where  $J_{\pm} = J_x \pm iJ_y$ are the raising and the lowering
operators for $J_0 = J_z$.

Schr\"odinger's equation written in the spherical coordinates as
\begin{equation}
\bigg(-\frac{1}{2mr^2}\frac{\partial}{\partial r} \bigg( r^2
\frac{\partial}{\partial r}\bigg)+ \frac{({\mathbf J}^2 -
\mu^2)}{2mr^2} \bigg)\Psi = E \Psi, \label{eq01}
\end{equation}
admits the separation of variables and, putting $\Psi=
R(r)Y(\theta, \varphi)$ into Eq. (\ref{eq01}), we get
\begin{eqnarray}
&&\left(-\frac{1}{2mr^2}\frac{d}{dr}\bigg(r^2\frac{d}{dr}\bigg) +
\frac{l(l
+ 1) - \mu^2}{2mr^2}\right)R(r) = ER(r), \nonumber\\
&&{\mathbf J^2}Y(\theta,\varphi) = l(l + 1)Y(\theta, \varphi).
\label{eq7a}
\end{eqnarray}

Starting from $J_0 Y_{\mu} =mY_{\mu}$ and writing
\begin{eqnarray*}
&&Y_{\mu} =e^{i\alpha\varphi}z^{\alpha/2}(1-z)^{\beta/2}F, \\
&&\alpha=m + \mu,\;\beta=m - \mu
\end{eqnarray*}
where $z=(1-\cos\theta)/2$, we obtain the resultant equation in
the standard form of the hypergeometric equation,
\begin{eqnarray}
z(1-z)\frac{d^2F}{dz^2} +\bigl(c-(a+b+1)z\bigr)\frac{dF}{dz}-abF=0
\label{hyp1}
\end{eqnarray}
where
\begin{eqnarray}
&&c = m+\mu+1,\; a+b = 2m +1,\nonumber\\
&&ab =(m-l)(l+m+1). \label{hyp2}
\end{eqnarray}

The hypergeometric function $F(a,b;c;z)$ diverges when $\Re
(c-b-a)\leq -1$, and it reduces to a polynomial of degree $n$  in
$z$ when $a$ or $b$ is equal to $-n, \;(n = 0,1,2, \dots)$. For
$a$ being negative integer we find that the corresponding solution
of Eq.(\ref{hyp1}) is of the form \cite{Ab,GAN}
\begin{eqnarray}
F=z^\delta{(1-z)}^\gamma p_n(z) \label{pol}
\end{eqnarray}
where $p_n(z)$ is a polynomial in $z$ of degree $n$.

Here we are looking for the regular solutions, like (\ref{pol}),
of the Schr\"odinger equation (\ref{eq7a}). The requirement of the
wave function being single valued force us to take $m +\mu$ as an
integer. The respective regular solution is given by
\begin{align}
&Y_{\mu}=
C_{lm\mu}\,e^{i(m+\mu)\varphi}z^{\alpha/2}(1-z)^{\beta/2}F(a,b;c;z),
\label{hyp3}\\
&\alpha=m + \mu,\;\beta=m - \mu,\; c=m+\mu+1 \nonumber
\label{hyp3}
\end{align}
where $C_{lm\mu}$ is the normalization and for the parameters $a$
and $b$ we have:
\begin{align*}
&a=-n,\; b=n +\alpha + \beta+1,\; {\rm if}\; \alpha=0,1,2, \dots ,\\
&a=n+1,\; b=-n- \alpha -\beta,\; {\rm if}\; \alpha=-1,-2, \dots .
\end{align*}

It follows that $F$ reduces to the Jacobi polynomials
$P_n^{(\alpha,\beta)}$ so that $Y_\mu$ takes the form (compare
with \cite{Wu2,Lyn})
\begin{align*}
&Y_l^{(\mu,n)}
=C_{ln\mu}\,e^{i\alpha\varphi}(1-u)^{|\alpha|/2}(1+u)^{|\beta|/2}
P_n^{(|\alpha|,|\beta|)}(u),
\end{align*}
$\alpha= l+ \mu -n, \;\beta=l- \mu -n$ and $l = m +n$. Since $m +
\mu$ is an integer we conclude that $l + \mu$ must be an integer
too.

The function $Y_{l}^{(\mu,n)}$ is a member of a family
$\{Y_{\kappa,l}^{(\mu,n)}\}$ such that
\begin{equation}
Y_{\kappa,l}^{(\mu,n)}= \mbox{e}^{-i2\kappa\mu\varphi}
Y_l^{(\mu,n)} \label{eq2_a}
\end{equation}
is a solution of the Schr\"odinger equation corresponding to the
vector potential
\[
{\mathbf A}^{\kappa} =\kappa{\mathbf A}_{S} + (1-\kappa){\mathbf
A}_{N}.
\]
The requirement $Y_{\kappa,l}^{(\mu,n)}$ being single valued
yields $2\kappa \mu$ being integer. Thus, for a given $\mu$ a
weight $\kappa$ is quantizied parameter in units of $\mu$.

The wave functions $Y_{\kappa,l}^{(\mu,n)}$ form a complete set of
orthonormal solutions that implies any solution
$\Psi(\theta,\varphi;\mu,\kappa)$ can be expanded as
\begin{equation}
\Psi = \sum_{ln} C_{ln}Y_{\kappa,l}^{(\mu,n) } ,\quad C_{ln}=
\langle Y_{\kappa,l}^{(\mu,n)}|\Psi \rangle. \label{eq3d}
\end{equation}

Similar consideration can be done for the vector potential
$\mathbf A_S$. In this case $(l-\mu)\in \mathbb Z$ and the
corresponding wave functions being
$Y_{\kappa,l}^{(-\mu,n)}=Y_{1-\kappa,l}^{(\mu,n)}$ form a complete
set of orthonormal solutions as well.

For $(l \pm \mu)$ and $2\kappa\mu$ all being integers we call the
functions $Y_{\kappa,l}^{(\pm \mu,n)}$ {\it weighted monopole
harmonics}. They are regular for the all allowed values of $l,n$
and $\mu$. When $n+\alpha$, $n+\beta$ and $n+\alpha+\beta$ all are
integers $\geq 0$ and $\kappa =0$ the weighted monopole harmonics
are reduced to the {\em monopole harmonics} introduced by  Wu and
Yang \cite{Wu2}, and the imposed here restrictions on the values
of $n,\alpha$ and $\beta$ yield the Dirac quantization condition.

\section{Nonunitary representations of the rotation group and
solution of Dirac's monopole problem}

It is known that the unitary representations of the rotation group
leads to Dirac's quantization condition, $2\mu \in \mathbb Z$
\cite{H,Zw_1,Str,Fi}. Thus, the unique way to avoid the Dirac's
rule is to consider nonunitray representations. In what follows,
assuming $\mu$ being arbitrary parameter, we are looking for
nonunitary representations of the rotation group relating to an
arbitrary magnetic charge \cite{W,Com2}.

For $l(l+1)$ being value of the Casimir operator
\begin{eqnarray}
C=J_0^2 +\frac{1}{2}(J_{-}J_{+}+J_{+}J_{-}). \label{Cas}
\end{eqnarray}
we denote the states by $|l,n\rangle, \; n=0,1,\dots,\infty$. For
the representations bounded below we obtain
\begin{eqnarray}
&&J _{+} |l,n\rangle = \sqrt{(2l+n)(n+1)}|l,n+1\rangle,
\label{J1} \\
&&J _{-} |l,n\rangle = -\sqrt{n (2l +n -1)}|l,n-1\rangle, \\
&&J _{0} |l,n\rangle =(l + n )|l,n\rangle. \label{J3}
\end{eqnarray}
The representation is characterized by the eigenvalue $l$ of the
highest-weight state:  $|l,0\rangle$ such that $J_{-} |l,0\rangle
= 0$ and $ J_{0} |l,0\rangle =l|l,0\rangle$. Comparing Eqs.
(\ref{J3}) with $J _{0} Y_l^{\mu,m} =m Y_l^{\mu,m}$ and
remembering that $m + \mu \in \mathbb Z$ (see Sec. 2) we conclude
that $l+\mu$ is an integer. Thus, the representation bounded below
also can be characterized by $l + \mu$ being integer. Taking into
account the restriction following from the Schr\"dinger equation:
$l(l+1)-\mu^2 \geq 0$, we find that the allowed values of $l$ are
\begin{eqnarray}
l= |\mu| + \{-(\mu +|\mu|) \} +k, \quad k = 0,1,2,\dots.
\label{eq_3a}
\end{eqnarray}

For the representation bounded above we have
\begin{eqnarray}
&&J _{+} |l,n\rangle = -\sqrt{n(2l+n-1))}|l,n-1\rangle,
\label{J_1a} \\
&&J _{-} |l,n\rangle = \sqrt{(n+1) (2l +n)}|l,n+1\rangle,  \\
&&J _{0} |l,n\rangle = -(l + n )|l,n\rangle. \label{J_3a}
\end{eqnarray}
This representation is characterized by the eigenvalue $-l$ of the
highest-weight state:  $|l,0\rangle$ such that $J_{+} |l,0\rangle
= 0$ and $ J_{0} |l,0\rangle = -l|l,0\rangle$. We found that in
this case $l-\mu$ is an integer and the allowed values of $l$ are
\begin{eqnarray}
&&l= |\mu| + \{\mu -|\mu| \} +k, \; k =  0,1,2,\dots.
\label{eq3_a}
\end{eqnarray}

The obtained representations can be realized in the space of
holomorphic functions of a complex variable $z$. Following
\cite{Jac1a} we assign a ``wave function" $\langle z|l,n\rangle$
by
\begin{align}
&(l+\mu)\Rightarrow \; \langle z|l,n\rangle = A z^n ,
\label{below} \\
&(l-\mu) \Rightarrow \; \langle z|l,n\rangle  =A z^{-2l-n},
\label{above}
\end{align}
where $A=\sqrt{\Gamma(2l+n)/\Gamma(n+1) \Gamma(2l-1)}$ is a
normalization, $\Gamma$ being the Gamma function. The monomials
(\ref{below}) and (\ref{above}) form the basis for the analytic
functions in the unit disc $D: |z| \leq 1$ and in $\widetilde D:
|z|\geq 1$ respectively.

The Lie algebra is realized by the differential operators:
\begin{align}
&J_{+}= z^2\partial_{z} + 2lz,\; J_{-}=-\partial_{z},\;
J_{0}= z\partial_{z} + l, \\
&[J_{+},J_{-}] = 2 J_{0}, \; [J_{0},J_{\pm}] = \pm J_{\pm},
\end{align}
and an arbitrary state of the representation is of the form
\begin{eqnarray}
f(z) = \sum_{n=0}^{\infty}f_n \langle z|l,n\rangle
\end{eqnarray}
The inner product of two holomorphic functions is defined as
follows:
\begin{align}
&(l+\mu)\Rightarrow \; \langle f|g\rangle = \frac{1}{2\pi i}\int_D
d\bar z
d z \frac{\bar f g}{(1-|z|^2)^{2-2l}}, \\
&(l-\mu)\Rightarrow \; \langle f|g\rangle = \frac{1}{2\pi
i}\int_{\widetilde D} d\bar z d z \frac{\bar f
g}{(|z|^2-1)^{2-2l}}.
\end{align}
With the introduced inner product the group representation is
infinite dimensional, irreducible and nonunitary.

Finite-dimensional representation arises when $l$ takes the
exceptional values $2l=p$ with $p$ being positive integer. In this
case the representation is unitary and  bounded  from above and
below. One has the standard selectional rules: $l= |\mu| +k, \; k
= 0,1,2,\dots,  \quad m= -l,\dots ,l,$ and the Dirac quantization
condition holds \cite{Wu2}.

Returning to the eigenvalues equations
\begin{eqnarray}
&&{\mathbf J^2}Y(z) = l(l + 1)Y(z), \\
\label{eq7_b} &&J_0 Y(z) = \pm (l+n) Y(z) \label{eq7_a}
\end{eqnarray}
we see that their solutions given by eigenfunctions
$Y_l^{(\mu,n)}(z)$ of  Eqs. (\ref{below}), (\ref{above}) satisfy
the Schr\"dinger equation (\ref{eq7a}). Introducing the wave
function as follows:  $\Psi(r,z)=R(r)Y(z) $, where $Y(z)$ is a
holomorphic function:
\begin{eqnarray}
Y(z) = \sum_{n=0}^{\infty}f_n \langle z|l,n\rangle
\end{eqnarray}
we obtain the solution of the monopole problem inside of the unit
disc and for an arbitrary monopole charge.

For $D_{\pm}$ being unit disc we relate $z\in D_{+}$ to the points
of the upper semi sphere $\Sigma_{+}$ via the stereographic
projection from the south pole and $z\in D_{-}$ to the points of
the lower semi-sphere $\Sigma_{-}$ via the stereographic
projection from the noth pole. Covering the two- sphere $S^2$ as
follows:  $S^2=D_{+}\cup D_{-}$, we have the solution of the
Schr\"odinger equation of the form
$\Psi(r,\theta,\varphi)=\Psi_{+}\cup\Psi_{-}$ for the whole
sphere. In the intersection $D_{+}\cap D_{-}$ the functions
$\Psi_{\pm}$ must satisfy the relation:  $\Psi_{+}=\Psi_{-}$.

\section{Gauge transformations and monopole charge
quantization}

Before proceeding let us note that with the representations
$(l\pm\mu)$ are related two string families:  $\{S^\kappa_{\mathbf
n}\}$ and $\{S^{\tilde\kappa}_{-\mathbf n}\}$. Their respective
vector potentials are
\begin{align}
&A^\kappa_{\mathbf n} = \kappa A_{\mathbf n} + (1 -
\kappa)A_{-\mathbf n}, \quad 2\kappa\mu \in {\mathbb Z},
\label{A_1} \\
&A^{\tilde\kappa}_{-\mathbf n}=\tilde\kappa A_{-\mathbf n}+ (1
-\tilde\kappa)A_{\mathbf n}, \quad 2\tilde\kappa\mu \in {\mathbb
Z}. \label{A_2}
\end{align}
and the change $S^\kappa_{\mathbf n} \rightarrow
S^{\tilde\kappa}_{-\mathbf n}$ is given by the following gauge
transformation:
\begin{eqnarray}
&&A^{\tilde\kappa}_{-\mathbf n} = A^{\kappa}_{\mathbf n} -
d\chi^{\gamma}_{\mathbf n}, \quad
 \tilde\kappa = 1-\kappa -\gamma,
\label{A_01}\\
&&d\chi^{\gamma}_{\mathbf n} = 2{\gamma} q\frac{(\mathbf r \times
\mathbf n)\cdot d\mathbf r}{r^2- (\mathbf n \cdot \mathbf r)^2},
\label{A_02}
\end{eqnarray}
$\chi_{\mathbf n}$ being polar angle in the plane orthogonal to
${\mathbf n}$.

We start with an observation that due to the string quantization
one has the equivalence relation: $2\kappa'\mu = 2\kappa\mu
\mod{\mathbb Z}$. Therefore, further we restrict ourselves by the
gauge transformations, that do not change the weight of the
string, $S^\kappa_{\mathbf n} \rightarrow S^{\kappa}_{\mathbf
n'}$. It produces the transformation $\bigl(\mathbf
A^{\kappa}_{\mathbf n}, \Psi^{\kappa}_{\mathbf n}
\bigr)\rightarrow \bigl(\mathbf A^{\kappa}_{\mathbf n'},
\Psi^{\kappa}_{\mathbf n'}\bigr) $ given by \cite{Sw_1,Br}
\begin{eqnarray}
&&\Psi^{\kappa}_{\mathbf n'}(\mathbf r) =
\exp\bigl(-ie\Phi^\kappa_{\mathbf n,\mathbf n'}(\mathbf
r)\bigr)\Psi^{\kappa}_{\mathbf n}(\mathbf r) \label{g_1a}
\end{eqnarray}
where the function $\Phi^\kappa_{\mathbf n,\mathbf n'}(\mathbf r)$
satisfies
\begin{eqnarray}
\mathbf A^{\kappa}_{\mathbf n}(\mathbf r) - \mathbf
A^{\kappa}_{\mathbf n'}(\mathbf r) = \nabla \Phi^\kappa_{\mathbf
n,\mathbf n'}(\mathbf r). \label{ag1a}
\end{eqnarray}

Let denote by $\mathbf n'= g\mathbf n , g\in\rm SO(3)$, the left
action of the rotation group induced by $S^\kappa_{\mathbf n}
\rightarrow S^\kappa_{\mathbf n'}$. From rotational symmetry of
the theory it follows immediately that an arbitrary gauge
transformation $\Psi^\kappa_{\mathbf n} \rightarrow
\Psi^\kappa_{\mathbf n'}$ can be undone by rotation $\mathbf r
\rightarrow  \mathbf r g$. Using this fact and adopting results of
\cite{Wu2,Jac,Jac1} we find that an arbitrary gauge transformation
$U_g$, producing the rotation of the string $S^\kappa_{\mathbf n}
\rightarrow S^\kappa_{\mathbf n'}$, is given by nonintegrable
phase factor,
\begin{align}
&U_g\Psi^\kappa_{\mathbf n}(\mathbf r) =
\exp(i\alpha^{\kappa}_1(\mathbf r, \mathbf
n;g))\Psi^\kappa_{\mathbf n} (\mathbf r),
\label{g_0}\\
&\alpha_1(\mathbf r;g)= e \int_{\mathbf
r}^{\mathbf r'} \mathbf A^\kappa_{\mathbf n}(\boldsymbol \xi)
\cdot d \boldsymbol \xi, \quad \mathbf r' =  \mathbf r g
\label{g_1c}
\end{align}
where the integration is performed along the geodesic
$\widehat{\mathbf  r \,\mathbf r'}\subset S^2$ and $\alpha_1$ is
the so-called first cochain \cite{Jac,Gr,F}. Actually, $U_g$ is an
operator of the parallel transport along the geodesics on the
two-dimensional sphere of the fixed radius $r$.

For a given cochain $\alpha_1$ a 2-cocycle $\alpha_2$ is defined
by
\begin{align}
&\alpha_{2}(\mathbf r ;g_1,g_2)=\delta\alpha_1=
\alpha_1(\mathbf r g_1;g_2)-\nonumber\\
& -\alpha_1(\mathbf r; g_1g_2)+ \alpha_1(\mathbf r ;g_1)
\end{align}
which satisfies $\delta\alpha_2 = 0$, and, for $\alpha_2$ being 2-cochain, a 3-cocycle $\alpha_3 = \delta\alpha_2$ is given by
\begin{align*}
&\alpha_3(\mathbf r;g_1,g_2,g_3))=
\alpha_2(\mathbf r g_1;g_2,g_3))
-\alpha_2(\mathbf r;g_1g_2,g_3))+ \\
&+\alpha_2(\mathbf r;g_1,g_2g_3))
-\alpha_2(\mathbf r;g_1,g_2).
\label{g_2e}
\end{align*}
Similarly one can introduce $n$-cocylce $\alpha_n(\mathbf r;g_1,g_2,\dots,g_n)$ \cite{Gr,F}.

Following \cite{Jac,Gr} let us define a 2-cochain, $\alpha_2$, by
\begin{equation}
\alpha_{2}(\mathbf r;g_1,g_2)= e \int_{\Sigma} \mathbf
B d{\mathbf s} = e\Phi\big|_\Sigma \label{g_2c}
\end{equation}
where $\Phi\big|_\Sigma$ is a magnetic flux through the geodesic
triangle $\Sigma \subset S^2$ spanned by $(\mathbf r,\mathbf
r {g_1},\mathbf r {g_1g_2})$.  Since $\mathbf B = \nabla \times \mathbf
A$ locally, but not globally then $\alpha_2$ is a 2-cochain and
not a 2-cocycle. Indeed, applying Stokes' theorem we get
\begin{eqnarray}
{\alpha}_2(\mathbf r;g_1,g_2)  = \delta\alpha_1(\mathbf
r;g_1,g_2) + \sigma (S^\kappa_{\mathbf n}, \Sigma)
\end{eqnarray}
where $\sigma =\int_{\Sigma}{\mathbf h}^\kappa_{\mathbf n}\cdot
d\mathbf s$ being contribution of the string is not zero if and
only if the string crosses $\Sigma$.

For computing $\sigma$ let us divide $R^3$ into $R^3_{+}$ and
$R^3_{-}$ by the plane passing through the origin of coordinates
and orthogonal to $\mathbf n$. Assuming that the string
$S^\kappa_{\mathbf n}$ crosses $\Sigma$ at a point $p_0$, we find
\begin{eqnarray}
\sigma = &\biggl\{
\begin{array}{l}
4\pi (1-\kappa)\mu,\quad  p_0 \in \Sigma\cap R^3_{-}\\
4\pi  \kappa\mu,\quad  p_0 \in \Sigma\cap R^3_{+}.
\end{array}
\label{strg_1a}
\end{eqnarray}
Since $2\kappa \mu$ is an integer, one has
\begin{eqnarray}
{\alpha}_2  = \delta\alpha_1 + 4\pi \mu\bigl|_{p_0} \mod 2\pi
{\mathbb Z} \label{strg_1}
\end{eqnarray}

Similar consideration of the gauge transformations
$S^\kappa_{\mathbf n} \rightarrow S^\kappa_{-\mathbf n}$, related
with the reflections,  yields
\begin{align*}
{\alpha}_2  = \delta\alpha_1 + 4\pi (1-2\kappa) \mu =
\delta\alpha_1 + 4\pi \mu \mod 2\pi \mathbb Z. \label{strg_2}
\end{align*}

Examining the composition of two operators $U_{g_1}$ and
$U_{g_2}$, we find that 2-cochain $\alpha_2$ occurs in its
composition law as follows:
\begin{equation}
U_{g_1}U_{g_2}\Psi^{\kappa}_{\mathbf n}(\mathbf r) =
\exp(i\alpha_2(\mathbf r ;g_1, g_2))U_{g_1g_2}\Psi^{\kappa}_{\mathbf n}(\mathbf r) \label{g_2a}
\end{equation}
where $g_1,g_2 \in{\rm SO(3)}$.

Consider now  three elements $g_1, g_2,g_3\in {\rm O(3)} $
producing the transformations $S^\kappa_{\mathbf n} \rightarrow
S^\kappa_{\mathbf n_1}, S^\kappa_{\mathbf n} \rightarrow
S^\kappa_{\mathbf n_2}, S^\kappa_{\mathbf n} \rightarrow
S^\kappa_{\mathbf n_3}$ respectively. Then the product of the
three operators is given by
\begin{align*}
 U_{g_1}\bigl(U_{g_2}U_{g_3}\bigr)\Psi^{\kappa}_{\mathbf n}(\mathbf r)
=\exp(i\alpha_3(\mathbf r ;g_1, g_2,g_3))
\bigl(U_{g_1}U_{g_2}\bigr)U_{g_3}\Psi^{\kappa}_{\mathbf n}(\mathbf r)
\end{align*}
where $\alpha_3$ is a {\it three cocycle}

From Eqs.(\ref{g_2c}) and (\ref{strg_1}) it follows $\alpha_3= 4\pi
\mu\; \mod 2\pi \mathbb Z$ if the monopole is enclosed by the
geodesic simplex with vertices $(\mathbf r,\mathbf r g_1,
\mathbf r g_1g_2,\mathbf r g_1g_2g_3)$ or zero otherwise \cite{Com1}.

We turn now to Eq.(\ref{g_2a}) and rewrite the product of the two
transformations as
\begin{equation}
U_{g_1}U_{g_2}\Psi^\kappa_{\mathbf n}(\mathbf r) =
U_{\varphi(g_1,g_2;\mathbf r)}
\Psi^\kappa_{\mathbf n}(\mathbf r) \label{g_3a}
\end{equation}
where $\varphi$ is defined by
\begin{eqnarray}
&&\varphi(g_1,g_2;\mathbf r)=
\alpha^\kappa_1(\mathbf r;g_1 g_2)+
\alpha_2(\mathbf r;g_1, g_2) =\nonumber\\
&&= \alpha_1(\mathbf r;g_1)+ \alpha_1(\mathbf r;g_2) +\sigma (S^\kappa_{\mathbf n}, \Sigma).
\label{lp1}
\end{eqnarray}

It is easy verify that the following identity of
quasiassociativity holds:
\begin{align}
\varphi(g_1, \varphi(g_2,g_3;\mathbf r);\mathbf r)
 =\varphi(\varphi(g_1,g_2;\mathbf r g_3),g_3 ;\mathbf r).
\label{g_3b}
\end{align}
We say that Eqs.(\ref{g_3a})-(\ref{g_3b}) define a {\it gauge
loop}. This is a special case of transformation quasigroup
introduced by Batalin \cite{Bat} and a 3-cocycle, being a `measure' of nonassociativity, can be related with an associator in theory of quasigroups and loops \cite{N1,N2,N3,Pfl,S1}.

The gauge loop is associated also with the loop QU(1)  defined as a
loop of multiplication by unimodular complex numbers
\cite{N1,N2,N3}:
\begin{align}
&{\rm e}^{i\alpha}\ast{\rm e}^{i\beta} = {\rm e}^{i\alpha\ast
\beta},
\label{q_4a} \\
&\alpha\ast \beta =\alpha+\beta + F(\alpha,\beta), \; F(\alpha,0)=
F(0,\beta)=0. \nonumber
\end{align}
Before proceeding notice that QU(1) is isomorphic to the group U(1) if
\begin{eqnarray}
&&F(\alpha, \beta)+F(\alpha\ast\beta,\gamma) - F(\beta,\gamma)-\nonumber \\
&&- F(\alpha, \beta\ast\gamma) =0 \mod 2\pi{\mathbb Z},
\label{as1}
\end{eqnarray}
that is a 2-cocycle condition $\delta \alpha_2 =0 \mod
2\pi{\mathbb Z}$.

Assuming QU(1) to be a local loop we define a respective gauge
loop over $S^2$ by
\begin{eqnarray}
&&U_{\alpha(\mathbf r)}\Psi^\kappa_{\mathbf
n}(\mathbf r)= \exp(i\alpha(\mathbf r))\Psi^\kappa_{\mathbf
n}(\mathbf r), \\
&&U_{\alpha(\mathbf r)}U_{\beta(\mathbf r)}\Psi^\kappa_{\mathbf
n}(\mathbf r) =U_{\alpha(\mathbf r)\ast\beta(\mathbf
r)}\Psi^\kappa_{\mathbf n} (\mathbf r). \label{lp2}
\end{eqnarray}
Here the operation $\alpha(\mathbf r)\ast \beta(\mathbf r)$ is
given by Eq.(\ref{q_4a}) with $F(\alpha,\beta;\mathbf r)$
determined as follows: $F= \sigma (S^\kappa_{\mathbf n},\Sigma)$
where the geodesic triangle $\Sigma \subset S^2$ is spanned by
$(\mathbf r, \mathbf r g_\alpha,\mathbf r g_\beta), \; g_\alpha,
g_\beta \in\rm SO(3)$. For computing $g_\alpha$ we employ the rotational symmetry of the theory. This implies that for a given
string $\mathbf S^\kappa_{\mathbf n}$ and gauge function $\alpha(\mathbf r)$ the following equation holds:
\begin{align}
&\Psi^{\kappa}_{\mathbf n}(\mathbf r') = \Psi^{\kappa}_{\mathbf
n'}(\mathbf r) = \exp\bigl(i\alpha(\mathbf r)\bigr)
\Psi^{\kappa}_{\mathbf n}(\mathbf r),
\label{g_1_c} \\
&\mathbf r'=  \mathbf r g_{\alpha}, \; \mathbf
n'= g_{\alpha}  \mathbf n , \; g_{\alpha}\in {\rm SO(3)}.
\nonumber
\end{align}
It should be considered as the equation for finding $g_\alpha$. Returning now to Eq. (\ref{lp2}) we see that the local loop QU(1) becomes the gauge loop defined by Eqs. (\ref{g_3a}), (\ref{lp1}).

\section{Discussion and concluding remarks}

We deduced a consistent pointlike monopole theory, with an
arbitrary magnetic charge, involving nonunitary representations of
the rotation group and making use of nonassociative QU(1) bundle over $S^2$, where QU(1) is the structure loop \cite{N1,N2,N3}. From our approach it follows a generalized quantization condition, $2\kappa\mu \in \mathbb Z$, that can be considered as quantization of the weight string instead of the monopole charge. In particular cases $\kappa =1$
and $\kappa=1/2$ it yields the Dirac and Schwinger selectional rules respectively.

At first sight our results are in contradiction with well known
topological and geometrical arguments in behalf of Dirac quantization rule \cite{Sw_1,Wu1,G3}. For the better understanding of the problem let us notice that known proofs are based on employing unitary finite-dimensional representations of the rotation group or classical fibre bundle theory. One can remove the effect of 3-cocycle imposing the Dirac quantization condition, however, this arises only from a realization of the monopole as U(1) bundle over $S^2$ \cite{Wu1,Wu2,Gr}. This implies that there exists the division of space into overlapping regions $\{U_i\}$ such that nonsingular vector potential can be defined and yields the correct monopole magnetic field in each region. On each intersection $U_i\cap U_j$ can be defined the transition functions $q_{ij} = e\Phi_{{\mathbf n}_i{\mathbf n}_j}$ such that $U_i\cap U_j\rightarrow \rm U(1)$.
On the triple overlap $U_i\cap U_j\cap U_k$ it holds
\begin{equation}
\exp(i(q_{ij}+q_{jk}+q_{ki})) =\exp(i4\pi\mu),
\end{equation}
and the consistency condition requires  $q_{ij}+q_{jk}+q_{ki} = 0
\mod2\pi \mathbb Z$. This gives $2\mu \in \mathbb Z$ and the Dirac
quantization condition appears again, now as a {\it necessary
condition to have a consistent U(1)-bundle over $S^2$}. Notice that it is consequence of the dynamics and not of the representation theory \cite{Gr}.

While the Jacobi identity holds for the generators of the rotation group \cite{Zw_1,H,Str} the situation with the translations in the background of the monopole is quite different. The difference has a topological nature and arises from the non-trivial toplogy of the orbit space. In the case of the rotations, the orbit space is just a two-dimensional sphere $S^2$. For the translations the orbit space is three-dimensional space $R^3$ with one point removed and its non-trivial topology provides the non-vanishing three-cocycle \cite{G1}. Thus, the Jacobi identity fails for the gauge invariant algebra of translations and for the finite translations $\{U_{\mathbf a}\}$ one has \cite{Jac,Gr}
\begin{align}
\bigl(U_{\mathbf a}U_{\mathbf b}\bigr)U_{\mathbf c}\Psi(\mathbf r) =\exp(i\alpha_3(\mathbf r ;\mathbf a, \mathbf b,\mathbf c)) U_{\mathbf a}\bigl(U_{\mathbf b}U_{\mathbf c}\bigr)\Psi(\mathbf r)
\label{as}.
\end{align}
For the Dirac quantization condition being satisfied one has $\alpha_3 = 0 \mod 2\pi \mathbb Z$, and (\ref{as}) provides an associative representation of the translations, in spite of the fact that the Jacobi identity continues to fail.

Since a conventional quantum mechanics deals with linear Hilbert space operators, the Dirac quantization rule is a necessary condition for the consistency of quantum mechanics in the presence of a monopole. Avoiding this condition forces us to go beyond the standard quantum mechanical approach and introduce a {\it nonassociative algebra of observables} \cite{Jac,Gr,Gr1,G1,G2}. Notice that in ordinary quantum mechanics the Schr\"odinger and Heisenberg pictures are equivalent, but the same is not true in a nonassociative quantum mechanics. Indeed, whilst the concept of the Hilbert space failed for nonassociative algebras, the Heisenberg approach could be still realized \cite{Adl,OK,K}. In a possible nonassociative quantum mechanics one must give up a conventional description of the quantum mechanics provided by Hilbert space concept and look for the generalization based on the Heisenberg approach and maybe only in terms of density matrix \cite{G2,OK,K}.


\begin{thebibliography}{99}
\bibitem {Com2} We follow the ideas of \cite{Jac1a} where the description
of infinite dimensional unitary representations has been done for
the group SU(1,1).
\bibitem {Com1} Similar analysis related with the modification of the
translation group in the presence of the magnetic monopole has
been done in \cite{Jac} (see also \cite{Gr,Gr1,G1,G2}).
\end{thebibliography}
\end{document}